\documentclass[prb,aps,twocolumn,amsmath,amssymb,floatfix,showpacs,
superscriptaddress]{revtex4}

\usepackage[dvips]{graphics}
\usepackage{color}
\definecolor{gold}{rgb}{0,0,0.75}

\begin{document}

\title{\textcolor{gold}{Conformation dependent magnetotransport in a 
single handed helical geometry}}

\author{Srilekha Saha}

\affiliation{Condensed Matter Physics Division, Saha Institute of Nuclear 
Physics, Sector-I, Block-AF, Bidhannagar, Kolkata-700 064, India}

\author{Santanu K. Maiti}

\email{santanu.maiti@isical.ac.in}

\affiliation{Physics and Applied Mathematics Unit, Indian Statistical
Institute, 203 Barackpore Trunk Road, Kolkata-700 108, India}

\author{S. N. Karmakar}

\affiliation{Condensed Matter Physics Division, Saha Institute of Nuclear 
Physics, Sector-I, Block-AF, Bidhannagar, Kolkata-700 064, India}

\begin{abstract}

Conformation dependent circular current is investigated in a single handed
helical geometry in presence of magnetic flux $\phi$ within a Hartree-Fock
mean field approach. The helical model is described by a set of non-planar
rings connected by some vertical bonds where each ring is formed by 
introducing a non-zero hopping between the atoms $a$ and $b$ as shown in
Fig.~\ref{helix}. By stretching and compressing the geometry, circular 
current can be regulated significantly and thus the system can be exploited
to design current controlled device at the nano-scale level. The proximity
effect between the atomic sites $a$ and $b$ is also discussed in detail
which exhibits interesting results.

\end{abstract}

\pacs{73.23.Ra, 73.23.-b, 71.27.+a}

\maketitle

\section{Introduction}

Inspection of circular current in low-dimensional conducting loops driven 
by magnetic flux $\phi$ has remained alive over past few decades since its
prediction~\cite{butt1} in $1983$ by B\"{u}ttiker {\em et al.} It is well 
known that an isolated conducting mesoscopic ring, threaded by an 
Aharonov-Bohm (AB) flux $\phi$ carries a net circulating current, 
the so-called {\em persistent current}~\cite{gefen}, which never decays 
with time. To achieve this non-vanishing circular current the prerequisites 
are~\cite{gefen}: (i) system size should be finite
and not too large and (ii) temperature ($T$) of the system should be
low enough. Actually these two i.e., system size and temperature
are interdependent in the sense that the average energy level spacing
$\Delta E$ must be greater than $k_BT$ ($k_B$ being the Boltzmann constant)
to obtain non-decaying circular current. With increasing system size
$\Delta E$ decreases which recommends low temperature limit, while 
for smaller systems non-zero circular current can be obtained even for a 
higher temperature region as for these systems $\Delta E$ becomes 
sufficiently large.

\begin{figure}[ht]
{\centering \resizebox*{6cm}{5.5cm}{\includegraphics{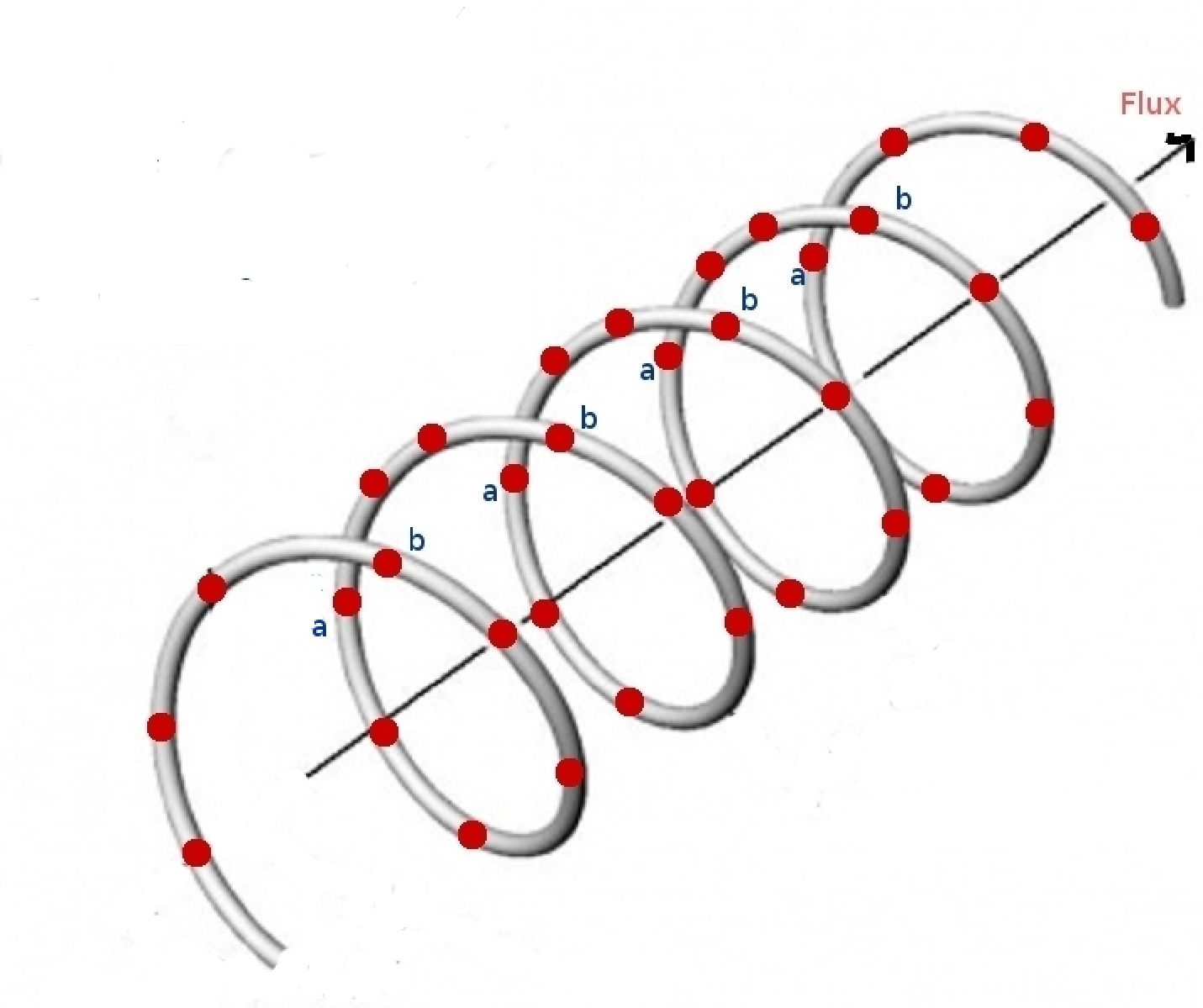}}\par}
\caption{(Color online). A single handed helical conductor represented by
a set of non-planar rings connected by some vertical bonds where each ring
is formed by introducing a finite hopping due to close proximity of the
atomic sites $a$ and $b$. A net circulating current is obtained in presence
of AB flux $\phi$ passing through centers of individual rings.}
\label{helix}
\end{figure}

Following the pioneering work done by B\"{u}ttiker and his co-workers,
many theoretical~\cite{ambe,schm1,schm2,bary,mailly1,mailly,smt,mont,ph,
sm1,sm2,sm3,sm4} and experimental~\cite{levy,jari,bir,chand,blu} groups 
have carried out systematic studies to explore interesting characteristic 
features of circular currents considering different loop geometries. 
In most of these cases simple ring-like conductors or array of rings or 
cylindrical conductors have been considered. Along with these few other 
geometries~\cite{g1,g2,g3,g4} have also been
taken into account to explore interesting patterns of circular current
in presence of Aharonov-Bohm flux $\phi$. In these works, persistent 
currents have been described in aspects of electron filling $N_e$, chemical
potential $\mu$ of the system, temperature $T$, electron-electron 
correlation $U$, electron-phonon interaction and to name a few. But, to 
the best of our knowledge, no one has addressed the issue of conformation
dependent circular current so far. In the present work we essentially 
focus on that particular issue by considering a single handed helical
geometry (Fig.~\ref{helix}). The helical model is illustrated by a set
of non-planar rings those are connected by some vertical bonds. Each of
these rings is formed by introducing a finite coupling between the atomic
sites $a$ and $b$ as shown in Fig.~\ref{helix}. The helical conductor 
exhibits a net circulating current in presence of AB flux $\phi$ passing 
through centers of the rings. The main motivation behind this work
is two-fold. (i) To investigate the interplay between conformational change 
and Hubbard correlation, which is still unaddressed, on circular current in
a helical geometry. (ii) The helical geometry has recently become truly
significant in understanding electron transport in several 
bio-molecules~\cite{bio1,bio2,bio3,bio4}.
Studying the behavior of circular current in presence of AB flux,
conducting nature of such helical-like geometries can be estimated up to
a certain level. We strongly believe that the present investigation may be
helpful in analyzing magnetotransport in several real as well as artificially
designed bio-molecules~\cite{bio1,bio2,bio3,bio4}.

Using a tight-binding (TB) framework we describe the model and compute
all the numerical results within a Hartree-Fock (HF) mean field 
level~\cite{mf1,mf2,mf3,mf4}. 
The results are impressive. (i) By stretching and compressing the helical
geometry we can control persistent current, keeping all the other physical
parameters unchanged. This is quite different from the conventional cases,
where current amplitude is regulated either by changing disorderness
or by means of electron filling or something else. Thus, our system can
be exploited to design conformation dependent current controlled device
at the nano-scale level. (ii) The other important observation is that the
geometry exhibits both $\phi_0/2$ and $\phi_0$ periodic persistent currents,
while conventional AB rings provide only $\phi_0$ periodic currents.

The rest of the work is organized as follows. In Section II, we illustrate 
the model and give a detailed theoretical description to calculate 
circulating current within a Hartree-Fock MF approach. The results 
are analyzed in Section III which includes the effects of compression and 
extension of the conductor, Hubbard correlation, electron filling, etc. 
Finally, In Section IV we summarize our findings and present the future 
perspective. 

\section{Model and Theoretical Framework}

\subsection{Model}

Let us begin by referring to Fig.~\ref{helix} where a single handed helical 
conductor in presence of magnetic flux $\phi$,
\begin{figure}[ht]
{\centering \resizebox*{6.5cm}{5cm}{\includegraphics{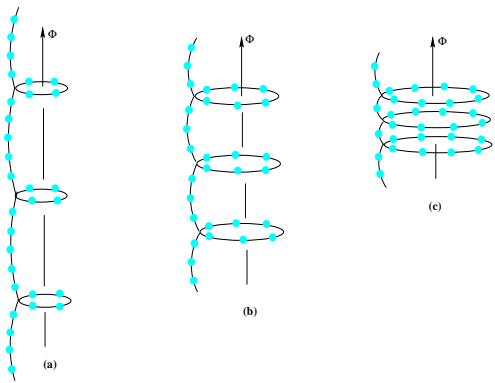}}\par}
\caption{(Color online). Schematic view of a helical conductor in 
presence of magnetic flux $\phi$ under (a) stretched, (b) undeformed 
and (c) compressed conditions.}
\label{combined}
\end{figure} 
measured in unit of the elementary flux-quantum $\phi_0$ ($=ch/e$), is 
given. Here we consider three different cases, viz, stretched, undeformed
and compressed, of the conductor depending on its configuration to
explore conformation dependent magnetotransport properties. These 
configurations are schematically shown in Figs.~\ref{combined}(a), (b) 
and (c), respectively, where the conformational change is described by 
increasing and/or decreasing the ring size and chain length such that the
total number of atomic sites $N$ of the helical conductor is fixed. It
is written as $N=R\,n_r + Wn_w^{\mbox{\tiny in}} + n_w^{\mbox{\tiny out1}}
+ n_w^{\mbox{\tiny out2}}$, where $R$ corresponds to the total number of
rings in which each ring contains $n_r$ atomic sites. The wire is divided
into three parts. One is called intermediate wire that connects two
adjacent rings and the other two are called outer wires. The parameter
$W$ represents total number of intermediate wires in the conductor where 
each of these wires holds $n_w^{\mbox{\tiny in}}$ atomic sites, while 
$n_w^{\mbox{\tiny out1}}$ and $n_w^{\mbox{\tiny out2}}$ give the atomic 
sites in the two outer wires, respectively.

The TB Hamiltonian of such a helical conductor can be written as a sum of
four terms like $H=H_R + H_W + H_W^{\mbox{\tiny out1}} + 
H_W^{\mbox{\tiny out2}}$, where they correspond to four different regions
of the interacting conductor. These terms are described as follows. The
Hamiltonian $H_R$ is given by $\sum_R H_{\mbox{\tiny ring}}$ where 
$H_{\mbox{\tiny ring}}$ describes identical rings, and, for any such rings
it becomes,
\begin{eqnarray}
H_{\mbox{\tiny ring}} &=& \sum_{\substack{i=1\\\sigma}}^{n_r} 
\epsilon_{i\sigma}^r c_{i\sigma}^{r\dagger} c_{i\sigma}^r + 
\sum_{\substack{i=1\\\sigma}}^{n_r-1} t\left[e^{i\theta} 
c_{i\sigma}^{r\dagger} c_{i+1\sigma}^r + h.c. \right] \nonumber \\
 & + & \lambda \sum_{\sigma} \left[e^{i\theta} c_{n_r\sigma}^{r\dagger} 
c_{1\sigma}^r + h.c. \right] + U \sum_{i=1}^{n_r} c_{i\uparrow}^{r\dagger}
c_{i\uparrow}^r c_{i\downarrow}^{r\dagger} c_{i\downarrow}^r. \nonumber \\
\label{eq1}
\end{eqnarray}
Here, $\epsilon_{i\sigma}^r$ is the site energy of an electron at $i$th 
site of spin $\sigma$ ($\uparrow$, $\downarrow$), $c_{i\sigma}^{r\dagger}$
and $c_{i\sigma}^r$ are the creation and annihilation operators, 
respectively. $t$ represents the nearest-neighbor hopping integral and 
$\theta$ ($=2\pi \phi/n_r\phi_0$) is the phase factor associated with 
magnetic flux $\phi$. $\lambda$ measures non-zero hopping between the 
atomic sites $a$ and $b$ (see Fig.~\ref{helix}) due to their close proximity
and $U$ gives the on-site Coulomb interaction.

In a similar fashion we express $H_W$ 
($=\sum_W H_{\mbox{\tiny wire}}^{\mbox{\tiny in}}$) which describes the 
Hamiltonian of all inner wires, and for any such wires it gets the form,
\begin{eqnarray}
H_{\mbox{\tiny wire}}^{\mbox{\tiny in}} &=& 
\sum_{\substack{i=1\\\sigma}}^{n_w^{\mbox{\tiny in}}} 
\epsilon_{i\sigma}^w c_{i\sigma}^{w\dagger} c_{i\sigma}^w + 
\sum_{\substack{i=1\\\sigma}}^{n_w^{\mbox{\tiny in}}-1} v\left[
c_{i\sigma}^{w\dagger} c_{i+1\sigma}^w + h.c. \right] \nonumber \\
 & & + U \sum_{i=1}^{n_w^{\mbox{\tiny in}}} c_{i\uparrow}^{w\dagger}
c_{i\uparrow}^w c_{i\downarrow}^{w\dagger} c_{i\downarrow}^w.
\label{eq2}
\end{eqnarray}
where, $v$ corresponds to the nearest-neighbor hopping integral and other 
parameters carry similar meaning like above.

Finally, to describe TB Hamiltonians $H_W^{\mbox{\tiny out1}}$ and
$H_W^{\mbox{\tiny out2}}$ for the two outer arms we use exactly similar
kind of TB Hamiltonian as given in Eq.~\ref{eq2}, except that in one
case $n_w^{\mbox{\tiny in}}$ is replaced by $n_w^{\mbox{\tiny out1}}$,
while in the other case $n_w^{\mbox{\tiny in}}$ is substituted by
$n_w^{\mbox{\tiny out2}}$. Both these outer and inner wires are coupled
to the rings via the same hopping integral $v$.

This is the complete description of the model quantum system considered in
this work, and, now we describe the Hartree Fock mean field scheme to 
evaluate ground state energy and finally to determine persistent current
as a function of flux $\phi$.

\subsection {Mean field approach}

To evaluate energy eigenvalues of the interacting helical conductor we
use generalized Hartree-Fock mean field approach~\cite{mf1,mf2,mf3,mf4} 
where we decouple the 
interacting Hamiltonian into the non-interacting ones such that one is
associated with up spin electrons and the other is related to down spin
electrons. In this prescription on-site energies get modified and they 
are expressed as:
\begin{subequations}
\begin{align}
\epsilon_{i\uparrow}^{r(w)\prime} &= \epsilon_{i\uparrow}^{r(w)} +
U \langle n_{i\downarrow}^{r(w)} \rangle \\
\epsilon_{i\downarrow}^{r(w)\prime} &= \epsilon_{i\downarrow}^{r(w)} +
U \langle n_{i\uparrow}^{r(w)} \rangle.
\end{align}
\end{subequations}
Here, $n_{i\sigma}^{r(w)}= c_{i\sigma}^{r(w)\dagger} c_{i\sigma}^{r(w)}$
is the number operator and $r$ and $w$ are used to relate the ring and wire,
respectively. With these modified on-site energies we can express the
Hamiltonians of the ring as well as connecting wires in the decoupled 
form under mean field approximation as follows.
\begin{eqnarray}
H_{\mbox{\tiny ring}}^{\mbox{\tiny MF}} &=& \sum_{i=1}^{n_r} 
\epsilon_{i\uparrow}^{r\prime} n_{i\uparrow}^r + \sum_{i=1}^{n_r-1} 
t\left[e^{i\theta} c_{i\uparrow}^{r\dagger} c_{i+1\uparrow}^r + 
h.c. \right] \nonumber \\
 & + & \lambda \left[e^{i\theta} c_{n_r\uparrow}^{r\dagger} 
c_{1\uparrow}^r + h.c. \right] \nonumber \\
 & + & \sum_{i=1}^{n_r} 
\epsilon_{i\downarrow}^{r\prime} n_{i\downarrow}^r + \sum_{i=1}^{n_r-1} 
t\left[e^{i\theta} c_{i\downarrow}^{r\dagger} c_{i+1\downarrow}^r + 
h.c. \right] \nonumber \\
 & + & \lambda \left[e^{i\theta} c_{n_r\downarrow}^{r\dagger} 
c_{1\downarrow}^r + h.c. \right] \nonumber \\
 & - & \sum_{i=1}^{n_r} U \langle n_{i\uparrow}^r \rangle 
\langle n_{i\downarrow}^r \rangle \nonumber \\
 &=& H_{\mbox{\tiny ring}\uparrow} + H_{\mbox{\tiny ring}\downarrow}
- \sum_{i=1}^{n_r} U \langle n_{i\uparrow}^r \rangle 
\langle n_{i\downarrow}^r \rangle
\label{eq4}
\end{eqnarray}
where $H_{\mbox{\tiny ring}\uparrow}$ describes the effective TB Hamiltonian
for up spin electrons and for down spin electrons it is expressed by 
$H_{\mbox{\tiny ring}\downarrow}$. The term $\sum_{i=1}^{n_r} U \langle 
n_{i\uparrow}^r \rangle \langle n_{i\downarrow}^r \rangle$ is the constant
term and it produces an energy shift. 

Similar to Eq.~\ref{eq4} we can write the TB Hamiltonian of the inner 
wires as,
\begin{eqnarray}
H_{\mbox{\tiny wire}}^{\mbox{\tiny in, MF}} &=& 
\sum_{i=1}^{n_w^{\mbox{\tiny in}}} \epsilon_{i\uparrow}^w n_{i\uparrow}^w + 
\sum_{i=1}^{n_w^{\mbox{\tiny in}}-1} v\left[c_{i\uparrow}^{w\dagger} 
c_{i+1\uparrow}^w + h.c. \right] \nonumber \\
& + & \sum_{i=1}^{n_w^{\mbox{\tiny in}}} \epsilon_{i\downarrow}^w 
n_{i\downarrow}^w + \sum_{i=1}^{n_w^{\mbox{\tiny in}}-1} 
v\left[c_{i\downarrow}^{w\dagger} c_{i+1\downarrow}^w + h.c. \right] 
\nonumber \\
 & - & \sum_{i=1}^{n_w^{\mbox{\tiny in}}} \langle n_{i\uparrow}^w \rangle
\langle n_{i\downarrow}^w \rangle \nonumber \\
&=& H_{\mbox{\tiny wire}\uparrow} + H_{\mbox{\tiny wire}\downarrow}
- \sum_{i=1}^{n_w^{\mbox{\tiny in}}} U \langle n_{i\uparrow}^w \rangle 
\langle n_{i\downarrow}^w \rangle
\label{eq5}
\end{eqnarray}
where different terms have their usual meanings.

Exact expression like Eq.~\ref{eq5} also goes to the two outer arms with 
appropriate upper limits of the summation index as described before.

Combining all these decoupled Hamiltonians for different sub-parts we 
can write the effective Hamiltonian for the full interacting conductor
in a general way, for the sake of simplification, as
\begin{eqnarray}
H^{\mbox{\tiny MF}} &=& H_{\uparrow} + H_{\downarrow} -\sum_i U 
\langle n_{i\uparrow} \rangle \langle n_{i\downarrow} \rangle.
\label{eq6}
\end{eqnarray}
Now we can go through self-consistent procedure with these decoupled
Hamiltonians setting initial guess values of $\langle N_{i\uparrow}\rangle$
and $\langle N_{i\downarrow}\rangle$. First we construct $H_{\uparrow}$ 
and $H_{\downarrow}$ considering these $\langle N_{i\uparrow}\rangle$ and
$\langle N_{i\downarrow}\rangle$ and then diagonalize these Hamiltonians 
to find new set of $\langle N_{i\uparrow}\rangle$ and
$\langle N_{i\downarrow}\rangle$. Again we repeat this procedure and 
continue it until a self-consistent solution is obtained. This is the
most crucial step for mean-field scheme, and therefore, special 
emphasis should be given to choose the initial guess values of 
$\langle N_{i\uparrow}\rangle$ and $\langle N_{i\downarrow}\rangle$.

Once we get the self-consistent solution, the ground state energy of the
system can be easily determined. At absolute zero temperature ($T=0\,$K)
it becomes,
\begin{equation}
E_g=\sum_p E_{p\uparrow} + \sum_p E_{p\downarrow} - \sum_i U 
\langle n_{i\uparrow} \rangle \langle n_{i\downarrow} \rangle
\label{eq7}
\end{equation}
where the summation is taken upto the Fermi energy $E_F$. $E_{p\uparrow}$'s
and $E_{p\downarrow}$'s are the eigenvalues of the sub-Hamiltonians for
up and down spin electrons, respectively.

Finally, we determine persistent current of the helical conductor, at 
absolute zero temperature, from the relation~\cite{gefen,pc1},
\begin{equation}
I(\phi)=-\frac{\partial E_g(\phi)}{\partial \phi}
\label{eq8}
\end{equation}
where, $E_g(\phi)$ is the ground state energy in presence of flux $\phi$
which is determined by summing over lowest $N_e$ energy levels. $N_e$ 
represents the electron filling.

\section{Numerical results and discussion}

Now, we present the results which are computed numerically based on the 
above theoretical prescription (discussed in Sec. II). The common 
parameters for our calculations, unless stated otherwise, are: 
$\epsilon_{i\uparrow}^r=\epsilon_{i\downarrow}^r=\epsilon_{i\uparrow}^w=
\epsilon_{i\downarrow}^w=0$, $t=1\,$eV, $v=1\,$eV and $T=0\,$K. Throughout
the analysis we measure the energy in unit of $t$ and current in unit of
$et/h$. Here, we essentially focus on the variation of conformation dependent
circular current in a helical conductor in presence of magnetic flux $\phi$.
Before addressing this issue, first we describe the nature of energy-flux
characteristics to make the present work a self-contained study.

\begin{figure}[ht]
{\centering \resizebox*{8.5cm}{9cm}{\includegraphics{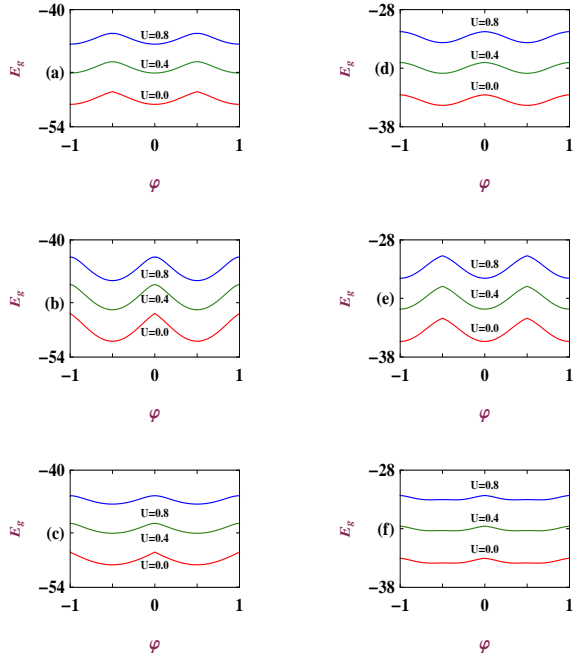}}\par}
\caption{(Color online). $E_g$-$\phi$ characteristics of a helical conductor
with $N=39$ (considering $R=4$ and $W=3$) for its different conformations 
where the left column corresponds to the half-filled band case i.e., 
$N_e=39$, while the other column represents $N_e=19$. The first row 
describes the undeformed conductor where we set 
$n_w^{\mbox{\tiny out1}}=n_w^{\mbox{\tiny out2}}=3$, $n_r=6$ and 
$n_w^{\mbox{\tiny in}}=3$. For the second row the results are shown for the
stretched conductor where we choose 
$n_w^{\mbox{\tiny out1}}=n_w^{\mbox{\tiny out2}}=4$, $n_r=4$ and
$n_w^{\mbox{\tiny in}}=5$. Finally, the last row represents the compressed
conductor with the parameters 
$n_w^{\mbox{\tiny out1}}=n_w^{\mbox{\tiny out2}}=2$, $n_r=8$ and
$n_w^{\mbox{\tiny in}}=1$. In each of these spectra $E_g$ is computed for
three different values of $U$ as shown by three distinct colored curves,
considering $\lambda=1\,$eV.}
\label{energy}
\end{figure}

\begin{figure}[ht] 
{\centering \resizebox*{8.7cm}{9cm}{\includegraphics{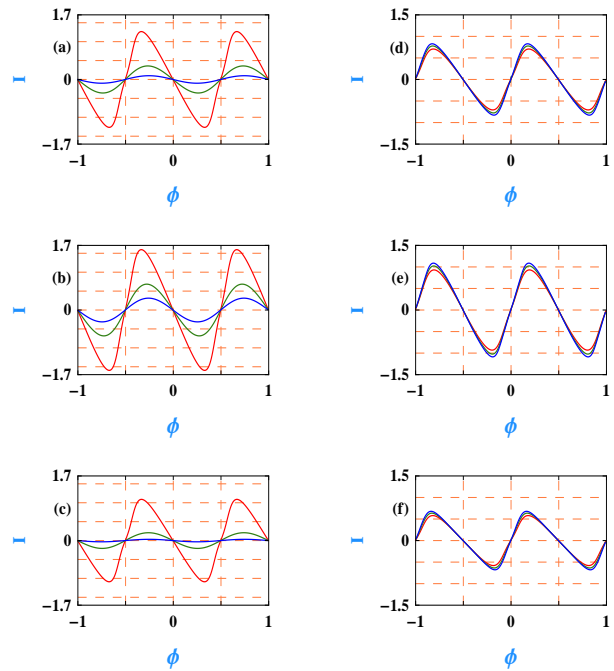}}\par}
\caption{(Color online). $I$-$\phi$ characteristics of a helical conductor
with $N=190$ (considering $R=4$ and $W=3$) and $\lambda=1\,$eV for its 
different conformations where the left column corresponds to the 
half-filled band case i.e., $N_e=190$, while the other column represents 
$N_e=95$. The three different rows represent the results for the three 
separate conformations as prescribed in Fig.~\ref{energy}. In the first 
row we set 
$n_w^{\mbox{\tiny out1}}=n_w^{\mbox{\tiny out2}}=14$, $n_r=30$ and 
$n_w^{\mbox{\tiny in}}=14$, while for the second row 
the parameters are
$n_w^{\mbox{\tiny out1}}=n_w^{\mbox{\tiny out2}}=18$, $n_r=22$ and
$n_w^{\mbox{\tiny in}}=22$. Finally, in the last row we consider
$n_w^{\mbox{\tiny out1}}=n_w^{\mbox{\tiny out2}}=10$, $n_r=38$ and
$n_w^{\mbox{\tiny in}}=6$. In each of these spectra $I$ is computed for
three different values of $U$, shown by three distinct colored curves,
where the red, green and blue lines in the left column correspond to
$U=0$, $0.5$ and $1$, respectively, and they are $0$, $1$ and $2$, 
respectively, for the right column.}
\label{current1}
\end{figure}

In Fig.~\ref{energy} we show the variation of ground state energy $E_g$ as
a function of magnetic flux $\phi$ for a helical conductor considering its 
three different configurations, viz, undeformed, stretched and compressed.
Two different band fillings are considered to compute the ground state 
energies and they are presented in two different columns. For the first
column we set $N_e=39$, while it is $19$ for the other column and in all 
these cases we determine $E_g$ considering three different values of $U$
to investigate the interplay between conformational change and on-site
Hubbard correlation. In this particular conductor we choose $N=39$ and
it is distributed among four rings ($R=4$) and three internal wires ($W=3$)
including two outer wires in appropriate numbers to get three distinct 
configurations of the helical conductor. Comparing the spectra given in 
Fig.~\ref{energy} it is observed that the ground state energy exhibits 
maximum variation with flux $\phi$ for the stretched configuration, 
irrespective of $U$ as well as band fillings ($2$nd row of Fig.~\ref{energy})
and it becomes much flatter as we move towards the compressed one. This
essentially leads to maximum circular current in the stretched case, while
lesser currents are obtained in other two cases, which are described later
in Fig.~\ref{current1}, since current is determined by taking the first
order derivative of $E_g$ with respect to flux $\phi$ (see Eq.~\ref{eq8}).
With this conformation dependent variation it is also important to note 
that the ground state energy significantly increases with the rise of $U$
and the slope of $E_g$-$\phi$ curve changes in a large and/or small scale
depending on the band filling, and it (change in slope) becomes much 
clearly seen from current-flux characteristics rather than $E_g$-$\phi$
spectra. All these ground state energies exhibit $\phi_0$ flux-quantum
periodicity.

In Fig.~\ref{current1} we present the variation of persistent current 
$I$ as a function of flux for a typical helical conductor with $N=190$
considering different values of $U$ for two distinct band fillings. In the
first column the currents are shown for $N_e=190$, while in the second
column we compute the currents setting $N_e=95$. The results are noteworthy.
It is observed that the current becomes maximum for the stretched geometry
and it gradually decreases as we increase the size of the ring i.e., when 
we move towards the compressed one. This behavior is independent of the
band filling as well as the Hubbard interaction strength. The signature of 
higher current for the stretched conductor and smaller current for the 
other two configurations can be clearly understood from the variation of 
energy-flux characteristics discussed above. In addition it is also 
important to note that for the half-filled band case ($N_e=190$) the 
current changes sharply with increasing $U$, while away from half-filling
($N-e=95$) it becomes less sensitive to $U$. In all these cases current
provides $\phi_0$ flux-quantum periodicity.
 
The results discussed so far, viz, $E_g$-$\phi$ and $I$-$\phi$ spectra,
are computed for some typical helical conductors where the number of
\begin{figure}[ht] 
{\centering \resizebox*{7cm}{7cm}{\includegraphics{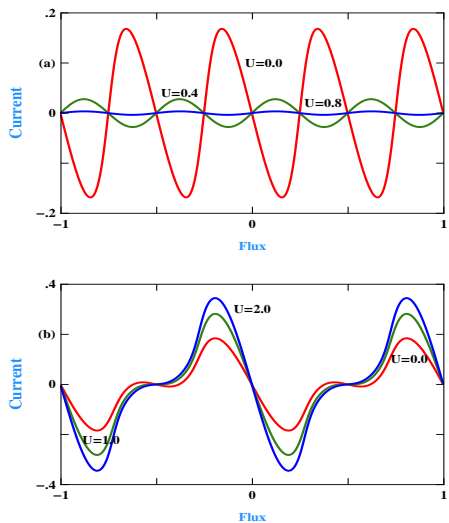}}\par}
\caption{(Color online). Current-flux characteristics for a helical
conductor with $N=190$ (considering $R=4$ and $W=3$) and $\lambda=1\,$eV 
in (a) half-filled ($N_e=190$) and (b) quarter-filled ($N_e=95$) cases. 
The other physical parameters are:
$n_w^{\mbox{\tiny out1}}=14$, $n_w^{\mbox{\tiny out2}}=15$, $n_r=29$ and
$n_w^{\mbox{\tiny in}}=15$.}
\label{current2}
\end{figure}
atomic sites $n_r$ in each ring is {\em even}. Now, to check if any anomalous
behavior is observed in current-flux characteristics for odd $n_r$, in 
Fig.~\ref{current2} we present the results of persistent currents 
considering helical conductor with $n_r=29$ (odd $n_r$). Two different cases
are shown, where in (a) currents are determined for $N_e=190$ (half-filled),
while in (b) they are computed when $N_e=95$ (quarter-filled). The results 
are quite interesting. For the half-filled band case current exhibits 
$\phi_0/2$ flux-quantum periodicity instead of the $\phi_0$ periodicity as
obtained in conventional cases. This typical behavior is only observed for 
the half-filled band case with odd ring size, since the current regains its
$\phi_0$ periodicity as long as the electron filling gets changed which is 
clearly observed by comparing the spectra given in Figs.~\ref{current2}(a)
and (b). From our extensive numerical analysis we find that this $\phi_0/2$
periodicity is the generic feature of persistent current for odd $n_r$ in
the half-filled band case, though its physical argument is not clear at this
stage and we hope it can be analyzed in our future work. Here, it is also
significant to state that the current amplitude sharply decreases with $U$
for the half-filled case, while quite comparable currents are obtained for
the other filling. In the limit of half-filling each atomic site is filled 
with an electron (up or down), and thus, it does not allow to hop an
electron from one site to the other due to repulsive nature of $U$ which
results reduced current with increasing $U$. While, for the system with less
than half-filling there is always an empty site, and therefore, electron
can hop into this empty site. The probability of electron hopping
increases with lower value of $U$, but it tends to decrease when $U$
exceeds the critical limit yielding a smaller current. This critical
value strongly depends on the system size and electron filling.

Finally, we discuss the proximity effect between the atomic sites $a$ and 
\begin{figure}[ht]
{\centering \resizebox*{6cm}{4cm}{\includegraphics{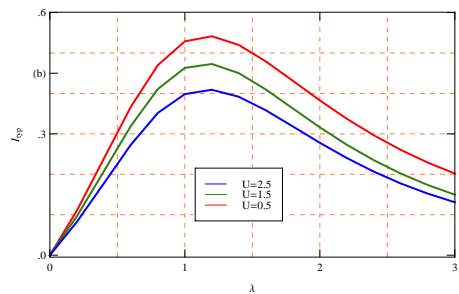}}\par}
\caption{(Color online). Current amplitude at a typical magnetic flux 
$\phi=0.35$ for a helical conductor with $N=190$ (considering $R=4$ and
$W=3$) as function of proximity integral $\lambda$ when $N_e$ is set at
$160$. The other physical parameters are:
$n_w^{\mbox{\tiny out1}}=n_w^{\mbox{\tiny out2}}=14$, $n_r=30$ and
$n_w^{\mbox{\tiny in}}=14$.}
\label{currlamb}
\end{figure}
$b$ on circular current. The results are presented in Fig.~\ref{currlamb},
where we show the variation of typical current amplitude as a function of
hopping integral $\lambda$ setting $\phi=0.35$. For all three different
values of $U$, we see that persistent current monotonically increases
and after reaching the maximum at $\lambda=1\,$eV, it eventually decreases
with $\lambda$. At this typical value of $\lambda$, the system behaves like
a perfect one since we set $t=v=1\,$eV in our theoretical formulation, 
which results a maximum current. On the other hand for all other cases
(when $\lambda \ne t$ and $v$) current decreases due to this anisotropy
in the hopping integral. This anisotropy is directly related to the 
closeness of the atomic sites $a$ and $b$, and thus, tuning the proximity
between these atomic sites we can regulate current amplitude which 
might be helpful in designing current controlled device at nano-scale 
level.

\section{Conclusion}

In the present work we investigate conformation dependent circular current
in a single handed helical conductor in presence of AB flux $\phi$. A 
tight-binding framework is given to illustrate the model where electronic
correlation is analyzed in the Hartree-Fock mean field level. Several 
important features are obtained from our numerical analysis those are 
summarized as follows. (i) Current amplitude can be controlled by 
stretching and compressing the geometry, keeping all other parameters
unchanged. (ii) Both $\phi_0$ and $\phi_0/2$ flux-quantum periodicities
are obtained in circular currents. The unconventional $\phi_0/2$ periodicity
is observed only for the half-filled band case when the conductor contains
rings with odd number of atomic sites. (iii) Circulating current can also
be regulated by tuning the coupling parameter between the atomic sites
$a$ and $b$. We believe that all the results studied here can be utilized
to investigate magneto-transport in several bio-molecular systems having
this helical like geometry.

In our analysis we compute the results for some typical parameter values of 
on-site energies, hopping integrals, system size, etc., but all these 
features remain unchanged for any other set of parameter values which
certainly demands an experimental verification towards this direction. 
The other valid approximation is the zero temperature limit, though finite
temperature extension of our analysis is a very simple task. The crucial 
point is that the physical properties studied above are not affected as
long as the average level spacing $\Delta E$ is greater than $k_BT$. For 
our geometry we can safely reach upto a sub-Kelvin temperature limit.

\end{document}